\documentclass[10pt,aps,pre,showpacs]{revtex4}
\usepackage{hyperref}
\usepackage{graphicx,subfigure}
\usepackage{amsmath,amsfonts,amssymb}

\newcommand{\f}{\frac}

\newcommand{\m}{\mathbf}
\newcommand{\oo}{\infty}
\newcommand{\intl}{\int\limits}

\begin{document}
\title{Asymmetry of the Hamiltonian and the Tolman's length}
\author{V. L. Kulinskii}
\affiliation{Department for Theoretical Physics, Odessa National
University, Dvoryanskaya 2, 65026 Odessa, Ukraine}
\begin{abstract}
Using the canonical transformation of the order parameter which
restores the Ising symmetry of the Hamiltonian we derive the
expression for the Tolman length as a sum of two terms. One of them
is the term generated by the fluctuations of the order parameter the
other one is due to the entropy. The leading singular behavior of
the Tolman length near the critical point is analyzed. The obtained
results are in correspondence with that of M.A. Anisimov, Phys. Rev.
Lett., \textbf{98} 035702 (2007).
\end{abstract}
 \maketitle
\section*{Introduction}\label{s1}
As is known (see e.g. \cite{widomrowlison}) the Tolman length can
be defined as the correction caused by the fact that the equimolar
surface does not coincide with the surface of tension for a small
droplet. The origin of such a difference is the asymmetry of the
phase coexistence in terms of the density variable
\cite{tolmanlengthmpfisher/prb/1984}. In its turn it is due to the
asymmetry of the Hamiltonian. In
\cite{tolmanlengthmpfisher/prb/1984} it was shown that within the
square gradient model the nonzeroth value of the Tolman length is
caused by the asymmetry of the density functional (Helmholtz free
energy). For the symmetrical models such as lattice gas the Tolman
length vanishes identically. Such a situation resembles the one
with the issue of the rectilinear diameter and its singularity
\cite{patpokr}. This effect is due to the asymmetry of the
Hamiltonian as a functional of the order parameter such as
density. The effects of asymmetry are consistently treated within
the approach based on the canonical form of the Hamiltonian
\cite{canformalokul/cmphukr/1997,canformkulinskii/jmolliq/2003,diamsingkulimalo/pre/2007}.
In such an approach both the linear \cite{patpokr,nicoll/pra/1981}
and nonlinear
\cite{fishmixdiam1/pre/2003,fisherdiam/chemphyslet/2005} mixing of
the thermodynamic fields are treated uniformly on the basis of the
isomorphism principle. In particular, the effects caused by the
asymmetry of the Hamiltonian are interpreted as the sequence of
improper choice of the order parameter. For proper, canonical
order parameter $\eta$ the Ising symmetry with respect to
transformation $\eta \to -\eta$ is restored. The main difference
between ``complete scaling`` approach and the proposed approach
the canonical form of the Hamiltonian lies in the fact that one
does not need to use the three scaling fields but work directly
with the Hamiltonian. In particular, it allows to make some
conclusions about the amplitudes of the singularities.

Recently, the question about critical behavior of Tolman length is
analyzed in \cite{tolmanlengthanis/prl/2007} within the ``complete
scaling`` approach. The leading singularities were obtained.

In this paper we demonstrate that the canonical formalism leads to
the essentially the same results obtained in
\cite{tolmanlengthanis/prl/2007} in addition giving the
possibility to relate the isomorphic variables with the
microscopic Hamiltonian. Basing on the expressions for the Tolman
length given in
\cite{tolmanlengthmpfisher/prb/1984,tolmanlengthblokhuis/jcp/2006}
we obtain the expression for the Tolman length which explicitly
demonstrates the role of the asymmetry of the Hamiltonian.

The structure of paper as follows. In Section~\ref{secnashort} we
give short outline of the procedure of the reduction of the
Hamiltonian to the canonical form. In Section~\ref{sectolmanlength}
we use the canonical transformation for structuring the Tolman
length.
\section{The reduction of the Hamiltonian to the canonical form}\label{secnashort}
We consider the case of 2-nd order phase transition. The typical
examples are Ising model and simple molecular liquids. For these
systems the Hamiltonian takes a form \cite{yukhgol}:
\begin{equation}
\label{h} \mathcal{H}[\rho(\mathbf{r})]=H_{ql}[\rho(\mathbf{r})]+
\int H_{loc}(\rho(\mathbf{r}))d\mathbf{r}\,,
\end{equation}
where
\begin{equation}\label{locham}
H_{loc}(\rho(\mathbf{r});\{a_{n}\})=
\sum\limits_{n=1}^{\infty}\frac{a_{n}}{n} \, \rho^{n}(\mathbf{r})\,
.
\end{equation}
For convenience we include $\beta=\frac{1}{k_{B}T}$ into the
Hamiltonian. In a case of simple liquids, for example, the
coefficients $a_{n}$ are definite functions of the chemical
potential $\mu$ (in fact the difference of the chemical potential
and its value $\mu(T)$ at the coexistence curve) and the temperature
$T$ if the nontrivial reference system is used \cite{yukhgol}.

Due to locality, for every point we can write:
\begin{equation}
\label{r} \rho(\mathbf{r})=F(\eta(\mathbf{r}))\,,
\end{equation}
then for the integrand in the partition function of the system we
have
\begin{equation}\label{can}
\exp\left(
-H^{(can)}_{loc}(\eta)\right)=\int\delta\left(\rho-F(\eta)\right)
\exp\left(-H^{(can)}_{loc}(\rho)\right)d\rho\,,
\end{equation}
where
\begin{equation}\label{hcan}
H^{(can)}_{loc}(\eta)=A_{1}\eta+\frac{A_{2}}{2}\eta^{2}+\frac{A_{4}}{4}\eta^{4}\,.
\end{equation}

Here we give the procedure of transforming the Hamiltonian to
canonical form using the ideas of the Catastrophe Theory
\cite{arnoldvargus1}. The link of the CF with the CT is the
transformation $\rho \to \eta(\rho)$ of the initial order
parameter. Similar idea of transformation of the variable reducing
the distribution to simpler (gaussian) form was used in
\cite{sornette/prep/2000}.

It is expedient to represent the local part of the fluctuational
Hamiltonian as the sum of even ``$(+)$`` and odd ``$(-)$`` parts:
\begin{equation}\label{hamloc}
H_{loc}(\rho(\mathbf{r}))=
\sum\limits_{n=1}^{\infty}\frac{a_{n}}{n} \, \rho^{n}(\mathbf{r})
=
H^{(+)}_{loc}(\rho(\mathbf{r}))+H^{(-)}_{loc}(\rho(\mathbf{r}))\,ю
\end{equation}
For the local part the Hamiltonian we can write:
\begin{equation}\label{trans1}
F[\tilde\eta] = G[\rho]\,,
\end{equation}
where
\begin{align*}
 F[\tilde\eta] = & \int\limits^{\tilde\eta}_{0}\exp\left(-H^{(can)}_{loc}(z)\right)dz \\
  G[\rho] =&
\int\limits^{\rho}_{0}\exp\left(-H_{loc}(z)\right)dz\,.
\end{align*}
The coefficients $A_{1}$ and $A_{2}$ of the canonical form are
determined basing on the equalities
\begin{align}\label{11transcond}
F[+\infty;A_1, A_2, A_4] =&\,  G[+\infty; \mu, T] \notag \\
F[-\infty;A_1, A_2, A_4] =&\,  G[-\infty; \mu, T]
\end{align}
which provide the bijectivity of the transformation given by
Eq.~\eqref{trans1}. To be specific we assumed that the coefficients
of the initial Hamiltonian depend on the ``laboratory`` variables
chemical potential $\mu$ and the temperature $T$. Two conditions
\eqref{11transcond} are not sufficient to fix three coefficients
$A_1, A_2, A_4$ as functions of the ``laboratory`` variables. The
coefficient $A_4$ can be fixed according to some additional
condition since it is assumed that $A_4 \ne 0$.

Let us describe how the coefficients of the canonical form relate
with the laboratory variables. Note that the Hamiltonian
\eqref{hamloc} is usually based on the mean-field equation of state.
Therefore neglecting the fluctuations the coefficients of the
canonical form can be found from the condition of invariance of the
CP locus within the local (mean-field) approximation (the
coefficient $a_3$ also vanishes at the CP due to stability reason
\cite{ll5}):
\begin{equation}\label{cposit}
A_{1}(\mu_c,T_c)=0\;,\,\,\, A_{2}(\mu_c,T_c)=0\quad
\Leftrightarrow\quad a_{1}(\mu_c,T_c)=0\;,\,\,\,
a_{2}(\mu_c,T_c)=0\,.
\end{equation}
These equations fix the value of $A_{4}$:
\begin{equation}
\int\limits_{-\infty}^{+\infty}
\exp\left(-\frac{A_{4}}{4}\eta^{4}\right)d\eta=
\int\limits_{-\infty}^{+\infty} \exp\left(-
H_{loc}(\rho;a_{1}=0,a_{2}=0)\right)d\rho\,,
\end{equation}
which gives
\begin{equation}\label{a4mf}
A^{(0)}_{4}=\frac{\pi^4}{\left(\Gamma\left(\frac{3}{4}\right)
\int\limits_{-\infty}^{+\infty}
\exp\left(-H_{loc}(\rho;a_{1}=0,a_{2}=0)\right)d\rho\right)^4}\,.
\end{equation}
But bearing in mind the account of fluctuations we do not need such
an approximation. Moreover the fluctuations change the locus of the
critical point. Taking into account the subsequent renormalization
of the Hamiltonian \eqref{hcan}, it is natural to put $A_4$ equal to
the renormalized interaction constant $u^*$ of the nongaussian fixed
point:
\begin{equation}\label{ac4}
  A_4 = u^*
\end{equation}
and use the representation $A_2 = r^* + \tau$, where $r^*$ is the
coordinate of the nongaussian fixed point $r^* =
-\f{\varepsilon}{6}\,\Lambda^2+o(\varepsilon)$ and $\Lambda$ is the
momentum cutoff \cite{ma}. In particular the critical point is
determined by:
\[A_{1}(\mu_c,T_c)=0\,, \quad A_{2}(\mu_c,T_c)= r^{*}\,,\]
Since the transformation \eqref{can} is smooth, the coefficients
$A_{i}$ are the smooth functions of the laboratory variables. In
particular at the coexistence curve for liquid-vapor transition we
have:
\begin{align}\label{a2a10}
 \left. A_{2}(\mu,T)\right|_{A_1=0} =&\,\, a\,\tau +o(\tau)\,,\\
a>0\,,\quad \tau
 =& \f{T-T_c}{T_c}\,.\notag
\end{align}
Thus the coefficients $A_1$ and $A_2$ of the canonical form can be
treated as the scaling fields. This is the approximation, because
the higher gradient terms are omitted. But since such terms do not
contribute to the renormalization of the local terms we can expect
that such an approximation incorporates the main thermodynamic
features of the interparticle interactions. Thus in variable $\eta$
the Hamiltonian functional takes the Landau-Ginzburg form:
\begin{equation}\label{hlg}
H_{LG}[\eta(\mathbf{r})]=\int
\left(A_{1}\eta+\frac{A_{2}}{2}\,\eta^{2}+
\frac{A_{4}}{4}\,\eta^{4} + \f{1}{2}\, (\nabla \eta)^2\right)
dV\,,
\end{equation}
where we scaled the square gradient term appropriately. Expanding
Eq.~\eqref{trans1} in a series we get:
\begin{equation}\label{cantranseries} \eta (\m{r}) =
\rho (\m{r}) + \f{1}{2}\Gamma_2\,\rho(\m{r})^2 +
\f{1}{3}\Gamma_3\,\rho(\m{r})^3 +\f{1}{4}\Gamma_4\,\rho(\m{r})^4 +
\ldots\,,
\end{equation}
where all coefficients are functions of the ``laboratory`` variables
(e.g. $\mu$ and $T$):
\begin{align}\label{gamma}
  \Gamma_2 =&\, a_1+A_1\,,\notag\\
  \Gamma_3 =&\,
  \f{1}{2}\left(\,a_{{2}}+\,{a_{{1}}}^{2}+\,A_{2}+A_{1}^{2}+3\,a_{1}
  A_{1}\right)\,,\\
    \Gamma_4 =&\,\f{1}{3}\,a_{3}+ \f{1}{2}\,a_{1}a_{2}+\f{1}{6}\,{a_{1}}^{3}+\frac
{7}{6}\,A_{1}A_{2}+\,{A _{1}}^{3}+\f{3}{2}\,a_{2}A_{1}+{\frac
{7}{6}}\,{ a_{1}}^{2}A_{1}+2\,a_{1}{A_{1}}^{2}+\,a_{1}A_{2} \,.
\notag
\end{align}

The approach proposed allows to treat correctly the effects caused
by the asymmetry of the Hamiltonian. Note that all information about
the asymmetry caused by the odd part of the local hamiltonian
$H_{loc}^{(-)}$ is represented by the linear term of the local part
of of the canonical form for the Hamiltonian with $A_1\ne 0$ and is
encoded also into $A_2$ and $A_4$.

The influence of asymmetry of the Hamiltonian is naturally to
analyse via the representation:
\begin{equation}\label{hrep}
  H_{loc}(\rho(\mathbf{r}))=\,H_{loc}^{(+)}(\rho(\mathbf{r}))+
  H_{loc}^{(-)}(\rho(\mathbf{r}))
\,,
\end{equation}
\begin{align}
\label{odev}
H_{loc}^{(+)}(\rho(\mathbf{r}))=&\,\frac{H_{loc}(\rho(\mathbf{r}))-%
H_{loc}(-\rho(\mathbf{r}))}{2}\,,\\
H_{loc}^{(-)}(\rho(\mathbf{r}))=&\,\frac{H_{loc}(\rho(\mathbf{r}))+%
H_{loc}(-\rho(\mathbf{r}))}{2}\,,
\end{align}
where superscripts $(+)$ and $(-)$ stand for even and odd components
of the function correspondingly. Assuming that the odd part of the
Hamiltonian is ``small`` in comparison with the even part, from
Eq.~\eqref{11transcond} we obtain:
\begin{equation}\label{a1}
  A_1 =\f{1}{c_1} \,\intl_{0}^{+\infty}\,H_{loc}^{(-)}(x)\,
  \exp\left(\,-H_{loc}^{(+)}(x)\,\right)
  dx + o\left(\,H_{loc}^{(-)}\,\right)\,,
\end{equation}
where
\begin{equation}\label{c1}
c_1=\frac{e^{\frac{A_2^2}{4 A_4}}
 \sqrt{\pi }}{2
   \sqrt{A_4}}\, \text{erfc}\left(\frac{A_2}{2 \sqrt{A_4}}\right)\,.
\end{equation}
Neglecting the fluctuational shift of the critical point this
result shows that the main contribution to $A_1$ comes from
$a_{5}$ (assuming that the coefficients $a_{2n+1}$ decrease with
$n$). In \cite{diamsingkulimalo/pre/2007} it was shown that the
singularity of the rectilinear diameter is shared by both
$\tau^{2\beta }$ and $\tau^{1-\alpha}$ anomalies, which are
generated by the asymmetrical part of the Hamiltonian. The result
of \cite{diamsingkulimalo/pre/2007} together with Eq.~\eqref{a1}
allows to relate the amplitudes of the $\tau^{2\beta}$ and
$\tau^{1-\alpha}$ singularities with the specificity of the
interparticle interaction of the system. In particular in
\cite{diamsingkulimalo/pre/2007} it was shown that these
amplitudes have opposite signs. This is in correspondence with the
results obtained in
\cite{fisherdiam/chemphyslet/2005,aniswangasym/prl/2006}. From
Eq.\eqref{a1} it follows that perturbatively this sign is
determined by the coefficient $a_5$. The value $\Gamma_2$ can be
considered as the asymmetry factor. Form Eqs.~\eqref{gamma},
\eqref{c1} it follows that such a factor is determined by the
asymmetrical part of the Hamiltonian.

Within the same approximation the coefficient $A_2$ is determined
implicitly as:
\begin{equation}\label{a2}
\f{1}{2}\sqrt{\f{A_2}{2 A_4}}\, e^{\frac{A_2^2}{8 A_4}}
\,K_{\frac{1}{4}}\left(\frac{A_2^2}{8 A_4}\right) =
\intl_{0}^{+\infty}\,  \exp\left(\,-H_{loc}^{(+)}(x)\,\right)
  dx
\end{equation}
where $K_{q}$ is the q-th order modified Bessel function of the 2-nd
kind.

The variables $A_1$ and $A_2$ unify the description of different
systems near the critical point. The relations Eqs.~\eqref{a1},
\eqref{a2} relate the coefficients of the canonical form
\eqref{hlg} with the laboratory variables. This dependence
determines the critical amplitudes for the specific system.

From Eq.~\eqref{cantranseries} it follows that for the average
value of the density we can write:
\begin{equation}\label{eqval}
\rho =\eta_{eq} + \eta_{asym} +\ldots
\end{equation}
where
\[\eta_{eq} = \left\langle\, \eta(\m{r})\,\right\rangle = \pm |A_2|^{\beta
}g_{\eta}\left(\,\f{A_1}{|A_2|^{{\beta+\gamma}}} \,\right) +
\text{Wegner corrections}
\] is the equilibrium value of the
canonical order parameter, %
and
\begin{equation}\label{etasym}
\eta_{asym} =    - \f{1}{2}\Gamma_2\left(\,\eta_{eq}^2\, +
s_{\eta}\right)\,,\quad
  s_{\eta} =
\left\langle\,\eta^2(\m{r})\,\right\rangle -
\left\langle\,\eta(\m{r})\,\right\rangle^2 =
\left|A_2\right|^{1-\alpha}\, l_{\eta}
\left(\,\f{A_1}{|A_2|^{{\beta+\gamma}}} \,\right)+\text{regular
terms}\,.
\end{equation}
is the part of the order parameter generated by the asymmetry of
the Hamiltonian. The representation \eqref{eqval} together with
Eq.~\eqref{etasym} gives the general basis for treating the
asymptotical properties of the physical quantities caused by the
asymmetry of the coexisting phases (e.g. liquid and vapor). In
particular, Eq.~\eqref{etasym} describes the rectilinear diameter
singularity \cite{diamsingkulimalo/pre/2007}.

Also it leads to the corresponding representation of the
dimensionless (isothermal) compressibility of the coexisting
phases into symmetric and asymmetric parts:
\begin{equation}\label{cancompress}
  \rho^2\,\chi_T =
\left.  \frac{\partial\, \rho }{\partial\, \mu }\right|_{T}  =
\left(\,\,\left. \frac{\partial\, A_{1}(\mu,T) }{\partial\, \mu
}\right|_{T}\,\frac{\partial\, }{\partial\, A_1} +\,\left.
\frac{\partial\, A_{2}(\mu,T) }{\partial\, \mu
}\right|_{T}\,\frac{\partial\, }{\partial\, A_2}\,\right)\,
\left(\,\eta_{eq}+\eta_{asym}\,\right)+ \ldots =
\tilde{\chi}_{sym}+\tilde{\chi}_{asym}
\end{equation}
analogous to that obtained in \cite{fishmixdiam1/pre/2003} (see
also \cite{tolmanlengthanis/prl/2007}). Despite similar ideology
(the nonlinear transformation of the order parameter) the
formalism of the canonical form of the Hamiltonian differs from
the ``complete scaling`` approach within the ``complete scaling``
 approach originally proposed in \cite{fishmixdiam1/pre/2003} to
 treat the singularity of the rectilinear diameter and to resolve
 the nature of Yang-Yang anomaly \cite{yang2/prl/1964}.
 The proposed approach is based
directly on the Hamiltonian. This leads to the prediction that
both $\eta^2_{eq}$ and $s_{\eta}$ contributions are generated by
the asymmetry of the Hamiltonian and proportional to asymmetry
factor $\Gamma_2$. While in ``complete scaling`` approach they are
in fact independent because of the phenomenological nature of the
hypothesis of complete scaling
\cite{fishmixdiam1/pre/2003,aniswangasym/prl/2006}. In addition
the proposed approach predicts that these two contributions have
opposite signs. This seems in correspondence with the estimates
found \cite{fishmixdiam1/pre/2003,fisherdiam/chemphyslet/2005} by
processing both the experimental data and model systems.

From Eq.~\eqref{cancompress} we see that in addition to the
standard $|\tau|^{-\gamma }$ singularity which is the same for
both coexisting phases we get the leading correction terms
$|\tau|^{\beta -\gamma}$ and $|\tau|^{1-\alpha - \beta -\gamma}$
which take opposite signs in these phases at the coexistence curve
due to $\eta_{asym}$. This result coincides with that of
\cite{tolmanlengthanis/prl/2007}. Since such terms according to
\cite{tolmanlengthanis/prl/2007} determine the critical behavior
of of the Tolman length we will consider the application of the
canonical formalism and the representation \eqref{eqval} to this
problem.
\section{Criticality of the Tolman length}\label{sectolmanlength}
According to \cite{tolmanlengthmpfisher/prb/1984} the Tolman length
as the coefficient of the asymptotic correction to the surface
tension of a drop of radius $R$ at $R\to \oo$ is as following:
\begin{equation}\label{tolmanlengthgrad}
  \delta_{\rho} = \f{\intl_{-\oo}^{+\oo} z\,\rho'(z) \,dz}{\intl_{-\oo}^{+\oo} \,\rho'(z)
  \,dz} - \f{\intl_{-\oo}^{+\oo} z\,\rho'^2(z) \,dz}{\intl_{-\oo}^{+\oo} \,\rho'^2(z)
  \,dz}
  \end{equation}
where $\rho(z)$ is the equilibrium density profile of interphase
coexistence. This profile is obtained basing on the minimization
square gradient functional (see e.g. \cite{widomrowlison}):
\begin{equation}\label{squaregrad}
  F[\rho(\m{r})] = \int \left(\, \f{m}{2}\,\left(\,\nabla \rho \,\right)^2 +
  f\left[\,\rho(\m{r})
  \,\right]
  \,\right)\,d\m{r}\,.
\end{equation}
which in fact is the LGH. In accordance with the result of previous
section we perform the local canonical transformation
\begin{equation}\label{cantranseries1}
\tilde{\rho}(\m{r}) = \eta(\m{r}) -
\f{1}{2}\Gamma_2\,\eta^2(\m{r}) + \ldots\,,
\end{equation}
where $\tilde{\rho}  = \rho(\m{r})/\rho_c-1$, which restore the
symmetry of the functional \eqref{squaregrad} in canonical
variable $\eta$.

Note that according to the definition, the spatial profile of the
canonical order parameter $\eta_{eq}(z)$ for two phase coexistence
is an odd function with respect
to the interphase boundary which is defined as the ``equi-$\eta$`` surface: %
\begin{equation}\label{etasym1}
\eta_{eq}(-z) = -\eta_{eq}(z)\,,%
\end{equation}
just like for any model with the even Landau-Ginsburg functional
\cite{tolmanlengthmpfisher/prb/1984}. The approach based on the
correlation functions \cite{tolmanlengthphillips/jcp/1985} gives
the same result. The second term in Eq.~\eqref{eqval} represents
the asymmetry effects
 \cite{canformalokul/cmphukr/1997,diamsingkulimalo/pre/2007}. In
 particular
 Eqs.~\eqref{eqval},\eqref{etasym} lead to the ``$\tau^{2\beta}$-`` and
``$\tau^{1-\alpha}$-`` anomalies of the the rectilinear diameter
\cite{diamsingkulimalo/pre/2007}. From Eq.~\eqref{eqval} it follows
that the phase coexistence profile of the density can be written as
follows:
\[\tilde {\rho}(z) = \eta_{eq}(z) + \eta_{asym}(z)+\ldots \]
with obvious Ising like properties
\begin{equation}\label{entsym}
\eta_{eq}(z) = -\eta_{eq}(-z)\,,\quad \eta_{asym}(z) =
\eta_{asym}(-z)\,,
\end{equation}
because of the symmetrical form of the Hamiltonian in canonical
variable $\eta$. Substituting this expression into
Eq.~\eqref{tolmanlengthgrad}
 we obtain:
\begin{equation}\label{tolmanlengthcan}
  \delta_{\rho} = -\Gamma_2 \,\delta_{can} + \ldots\,,
  \end{equation}
  where
\begin{equation}\label{tolmanlengthcan12}
\delta_{can} = \delta_{\eta} + \delta_{s}\,.
\end{equation}
Thus the amplitude value of the Tolman length is governed by the
value of $\Gamma_2$, which can be either positive or negative
depending on the details of the microscopic interaction. Since
$\eta_{eq} \propto |\tau |^{\beta }$ and $\eta_{asym} \propto |\tau
|^{2\beta }$ to the leading order we can write:
\begin{equation}\label{tolmanlengthcan22}
\delta_{\eta}= \f{1}{2}\,\f{\intl_{-\oo}^{+\oo}\,
z\,d\eta_{eq}^2(z)}{\intl_{-\oo}^{+\oo} \,d\eta_{eq}(z)} -
2\,\f{\intl_{-\oo}^{+\oo}\, z\,\eta_{eq}(z)\eta_{eq}'^2(z)
\,dz}{\intl_{-\oo}^{+\oo} \,\eta_{eq}'^2(z)
  \,dz}\,,\quad
\delta_{s}=\f{1}{2}\, \f{\intl_{-\oo}^{+\oo}\,
z\,d\,s_{\eta}(z)}{\intl_{-\oo}^{+\oo} \,d\eta_{eq}(z)} -
\f{\intl_{-\oo}^{+\oo}\, z\,s_{\eta}'(z) \,d\eta_{eq}
(z)}{\intl_{-\oo}^{+\oo} \,\eta_{eq}'^2(z)
  \,dz}
  \end{equation}
where $\eta(z)$ and $s_{\eta}(z)$ are the equilibrium profiles of
the canonical order parameter and the entropy correspondingly.

Further we assume that the density profile varies over the
correlation length $\xi$ as the only relevant characteristic
spatial scale near the critical point. Below we give the ground to
such an assumption using the rigorous thermodynamic expression for
$\delta$ obtained in \cite{tolmanlengthblokhuis/jcp/2006}. Then
simple scaling consideration shows that the obtained contributions
to Tolman length have the following leading singular behavior
\[\delta_{\eta} \propto \tau^{\beta
 -\nu}\,,\quad \delta_{s} \propto \tau^{1-\alpha -\beta
 -\nu}\,.\] This is exactly the result obtained in
 \cite{tolmanlengthanis/prl/2007}.
 The expression \eqref{tolmanlengthcan} for
 the Tolman length allows to give the ground for the approximate expression
\begin{equation}\label{deltanis}
  \delta \simeq -\xi \,\f{\rho_{d} -1 }{\Delta \rho}\,,
\end{equation}
proposed in \cite{tolmanlengthanis/prl/2007} basing on the
asymptotic critical behavior. Here $\rho_{d} = \f{\rho_{liq} +
\rho_{gas}}{2\rho_c}$ is the rectilinear diameter and $\xi$ is the
correlation length. From the point of view of Eq.~\eqref{eqval} it
is definitely right qualitatively. In order to obtain the critical
amplitudes for the behavior of the $\delta$ we use the rigorous
thermodynamic expression for the Tolman length given earlier in
\cite{tolmanlengthmpfisher/prb/1984,tolmanlengthbedauxblok/molphys/1993}
and recently represented in ``compressibility form`` in
\cite{tolmanlengthblokhuis/jcp/2006}:
\begin{equation}\label{tolmankenghtblokhuis}
\delta \approx -\sigma_{\infty}\,\f{\Delta\left(\,
\left.\frac{\partial\, \rho}{\partial\,
\mu}\right|_{T}\,\right)}{\left(\,\Delta \rho\,\right)^2}\,.
\end{equation}
Here $\sigma_{\infty}$ is the surface tension of planar interface.
Substituting Eq.~\eqref{eqval} and Eq.~\eqref{cancompress} into
Eq.~\eqref{tolmankenghtblokhuis} we obtain:
\begin{equation}\label{tolmanlengthtermodcan}
\delta \approx
2\f{\sigma_{\infty}}{\rho_c}\f{\tilde{\chi}_{asym}}{\left(\,\Delta
\tilde{\rho}\,\right)^2} = -\f{\sigma_{\infty}}{\rho_c}\,
\f{\left.\frac{\partial\, \eta_{asym}}{\partial\,
\mu}\right|_{T}}{4\eta_{eq}^2}\,.
\end{equation}
So that to the leading order we have:
\begin{equation}\label{tolmanlengthtermodcan1}
\delta \approx \f{\sigma_{\infty}}{\rho_c}\,
\left.\frac{\partial\, A_1}{\partial\,
\mu}\right|_{T}\,\Gamma_2\left(\,\f{g^{'}_{\eta}(0)}{g_{\eta}(0)}\,
|A_2|^{-\beta-\gamma}+
\f{l^{'}_{\eta}(0)}{g^2_{\eta}(0)}|A_2|^{-1-\beta}\,\right)+\ldots
\end{equation}

Both expressions for Tolman length Eq.~\eqref{tolmanlengthgrad}
and Eq.~\eqref{tolmankenghtblokhuis} give the same asymptotic
behavior provided that $\sigma \propto \xi^{-2} \propto
|A_2|^{2\nu} $ thus proving the assumption that the correlation
length is the characteristic scale for the spatial profile of the
density and the surface tension as the specific thermodynamic
potential of the interphase surface. From
Eq.~\eqref{tolmanlengthtermodcan1} it follows that the amplitude
of the Tolman length is determined by the asymmetry factor
$\Gamma_2$.

\section*{Conclusion}
In present paper within the canonical approach (see
\cite{canformalokul/cmphukr/1997,canformkulinskii/jmolliq/2003})
we show that the non zeroth value of the Tolman length is the
effect of the asymmetry of the Hamiltonian of the system in
density variable. Performing the transformation to the canonical
order parameter for which symmetry of the Hamiltonian is restored
we derive the invariant representation of the Tolman length in
terms of the profiles of canonical order parameter $\eta(z)$ and
canonical entropy $s(z)$. Such a representation allows to analyse
the asymptotic behavior of the Tolman length. The leading singular
terms are generated by the two above mentioned contributions and
proportional
 to $\propto \tau^{\beta -\nu }$ and $\propto \tau^{1-\alpha -\beta -\nu
}$ correspondingly. This is in correspondence with the results of
\cite{tolmanlengthanis/prl/2007} obtained within the ``complete
scaling`` approach of \cite{fishmixdiam1/pre/2003}. In fact the
qualitative representation \eqref{deltanis} shows that the nature
of the singularity of the Tolman length is determined by the
singularity of the rectilinear diameter $\rho_{d}$.
\begin{acknowledgements}
The author thanks to E.M. Blokhuis for valuable discussions of the
results.
\end{acknowledgements}


\end{document}